\documentclass[pra]{revtex4}
\usepackage{amsmath}
\usepackage{graphicx}
\begin{document}
\title{Aggregation Dynamics Using Phase Wave Signals and Branching Patterns}
\author{Hidetsugu Sakaguchi and Takuma Kusagaki}
\affiliation{Department of Applied Science for Electronics and Materials,
Interdisciplinary Graduate School of Engineering Sciences, Kyushu
University, Kasuga, Fukuoka 816-8580, Japan}
\begin{abstract}
The aggregation dynamics of slime mold is studied using coupled equations of phase $\phi$ and cell concentration $n$. Phase waves work as tactic signals for aggregation. Branching structures appear during the aggregation. A stationary branching pattern appears like a river network, if cells are uniformly supplied into the system.
\end{abstract}
\maketitle
\section{Introduction}
Target and spiral patterns are characteristic patterns in excitable and oscillatory media such as the Belousov-Zhabotinsky reaction~\cite{Winfree,Kuramoto}. Similar target and spiral patterns are also observed in the aggregation of the slime mold Dictyostelium discoideum~\cite{Gross}. The cAMP concentration exhibits limit-cycle oscillation in the slime mold. Martiel and Goldbeter proposed a mathematical model of the limit-cycle oscillation of cAMP~\cite{Martiel}, and Tyson et al. studied spiral patterns using the model of Martiel and Goldbeter~\cite{Tyson}. The cells aggregate to the center of the spiral using the cAMP signals. In the aggregation process, the cells take a characteristic branching structure. Keller and Segel proposed a model of the aggregation dynamics~\cite{Keller}. The model equations are coupled equations of the concentrations of cells and cAMP. 
 The aggregation toward the center of the spiral is not well explained in the simple model, because the cAMP concentration changes between high and low values in  the spiral wave. Van Oss et al. investigated  coupled equations for the cell concentration and cAMP waves, and numerically reproduced the aggregation toward the center of a spiral and the formation of branching patterns~\cite{Oss}. 
H\"ofer and Maini proposed an analytical model of streaming instability to explain branching patterns~\cite{Maini}. 
In previous papers, we proposed model equations for the aggregation dynamics using phase wave information~\cite{Sakaguchi1, Sakaguchi2}. The model equations are coupled equations of the cell concentration and the phase of cAMP oscillation. This is a modified model of the Keller-Segel model. That is, the concentration of cAMP in the Keller--Segel model is replaced with the phase of cAMP oscillation.  Since the phase waves do not decay in space, the signals for the aggregation can propagate for long distances effectively. Gregor et al. studied collective behaviors of the social amoebae~\cite{Gregor}. They found that several local aggregates are created initially. The competition occurs among local aggregates, and the number of local aggregates decreases over time, and finally only a dominant aggregate survives. Our model could reproduce the competitive aggregation dynamics. In the previous numerical simulation, however, each aggregation cluster takes a circular form and no characteristic branching patterns were observed clearly. In this study, we investigate branching patterns in the same model. 

\section{Two-dimensional Model for Aggregation and Branching Patterns}
The two-dimensional model equations are written as 
\begin{eqnarray}  
\frac{\partial\phi}{\partial t}&=&\frac{n}{1+\gamma n}+\nu\nabla^2 \phi+g(\nabla \phi)^2,\\
\frac{\partial n}{\partial t}&=&D\nabla^2n-c\left \{\frac{\partial}{\partial x}\left (\frac{\partial \phi}{\partial x}n\right )+\frac{\partial}{\partial y}\left (\frac{\partial \phi}{\partial y}n\right )\right \}, \label{2dmodel}
\end{eqnarray} 
where $\phi$ is the phase of the oscillation of the chemotactic factor such as cAMP, $n$ is the cell concentration, and $\gamma, \nu,g,D$, and $c$ are parameters. 
It is shown that the essential control parameters are $\gamma$, $D/\nu$, and $c\nu/g$ by the rescaling of the space and $\phi$.  
This model equation is a phenomenological model for understanding the aggregation dynamics.  The first term in Eq.~(1) implies that the frequency of the oscillation increases with the cell concentration, and the term including $\gamma$ represents the saturation effect at high concentrations. This type of sigmoidal behavior was observed in experiments~\cite{Gregor}. When $\gamma=0$, the collapse or the divergence of the local cell concentration often occurs in two dimensions.  
The second term in Eq.~(1) expresses the phase diffusion and the third term is a nonlinear term that appears by the standard phase reduction method~\cite{Kuramoto}. The first term in Eq.~(2) represents the diffusion of cells. The second term in Eq.~(2) denotes the chemotaxis using the phase wave information, that is, the cells are assumed to move toward a source of phase waves in proportion to the phase gradient $(\partial \phi/\partial x,\partial \phi/\partial y)$. The total number of cells $N=\int\int n(x,y,t)dxdy$ is conserved in the time evolution of Eq.~(2). 
The chemotaxis using the phase wave signals is not directly found experimentally. However, in the model by Van Oss et al., the cells are assumed to respond to the cAMP concentration only in increasing phases of cAMP waves.  Recently, this type of rectified directional response has been experimentally shown~\cite{Sawai}. If the temporal average is taken over one period of oscillation, cells move toward the source of cAMP waves on average. The second term in Eq.~(2) expresses this type of chemotaxis, because the temporal average over one period of oscillation is taken in the derivation of a phase equation. Our model is simpler than that of Van Oss et al., and exact solutions can be obtained in one dimension~\cite{Sakaguchi1}. 
\begin{figure}
\begin{center}
\includegraphics[height=3.5cm]{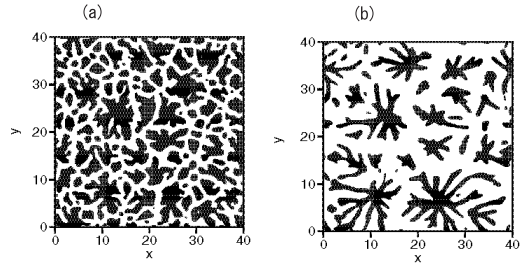}
\end{center}
\caption{Snapshot patterns of cell concentration at (a) $t=1000$, (b) 3000 for $\gamma=1, \nu=0.01,g=0.025,D=0.02$, and $c=0.0005$. The initial condition is $n(x,y)=0.4+r(x,y)$, where $r(x,y)$ is a random number between -0.01 and 0.01 and $\phi=0$. }
\label{f1}
\end{figure}
\begin{figure}
\begin{center}
\includegraphics[height=3.5cm]{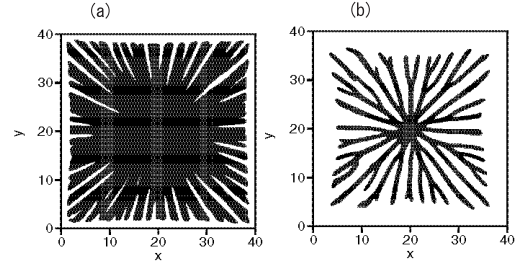}
\end{center}
\caption{Snapshot patterns of cell concentration at (a) $t=1000$, (b) 3000 for $\gamma=1, \nu=0.01,g=0.025,D=0.02$, and $c=0.0005$. The initial condition is slightly different from that shown in Fig.~1, that is,  $\phi=0$, $n(x,y)=1$ in the central region, and $n(x,y)=0.4+r(x,y)$, where $r(x,y)$ is a random number between -0.01 and 0.01 in the outer region.}
\label{f2}
\end{figure}
\begin{figure}
\begin{center}
\includegraphics[height=3.5cm]{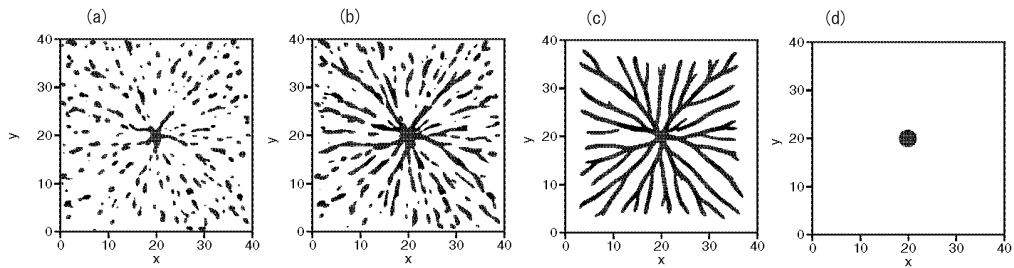}
\end{center}
\caption{Snapshot patterns of cell concentration at (a) $t=500$, (b) $t=750$, (c) 2500, and (d) 10000 for $\gamma=1, \nu=0.01,g=0.025,D=0.02$, and $c=0.0005$. As an initial condition, $n(x,y)$ takes 0.1 or 0.9 randomly with a probability of 0.8 or 0.2, and $n(x,y)$ is set to 1 at $t=0$ in the central region.}
\label{f3}
\end{figure}
\begin{figure}
\begin{center}
\includegraphics[height=3.5cm]{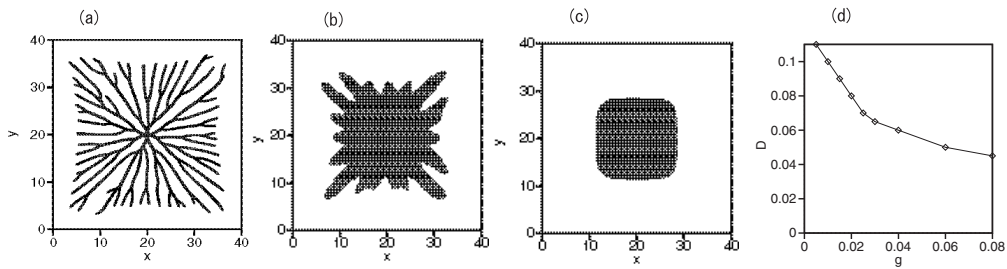}
\end{center}
\caption{Snapshot patterns of cell concentration at (a) $t=3000$ for $D=0.01$, (b) $t=5000$ for $D=0.06$, and (c) at $t=10000$ for $D=0.2$.  The initial condition is the same as that shown in Fig~.2. The other parameters are $\gamma=1,\nu=0.01, g=0.025$, and $c=0.0005$. (d) Transition line in a parameter space of $(g,D)$ for $\gamma=1,\nu=0.01$, and $c=0.0005$ above which branching patterns do not appear.}
\label{f4}
\end{figure}

There is a uniform solution of $\phi(t)=\phi(0)+n_0t/(1+\gamma n_0)$ and $n=n_0$ to Eqs.~(1) and (2). However, the solution is unstable for the perturbations $\delta\phi=\phi_{k_x,k_y}e^{ik_xx+ik_yy+\lambda_kt}$, and $\delta n=n_{k_x,k_y}e^{ik_xx+ik_yy+\lambda_kt}$ for a small $k=\sqrt{k_x^2+k_y^2}$, because $\lambda_k=\{-(D+\nu)k^2+\sqrt{(D-\nu)^2k^4+4cn_0k^2/(1+\gamma n_0)^2}\}/2>0$ for $k<\sqrt{cn_0/(D\nu)}/(1+\gamma n_0)$. 
We have performed numerical simulations for various parameter sets, but we show  numerical results only for $\gamma=1,\nu=0.01$, and $c=0.0005$. The system size is assumed to be $L\times L=40\times 40$, and periodic boundary conditions are imposed. The two-dimensional space is discretized with $\Delta x=\Delta y=0.1$, and the Heun method of the time-step $\Delta t=0.05$ is used for numerical simulation. Figures 1(a) and 1(b) show two snapshots of cell concentration where $n>0.4$ at (a) $t=1000$ and (b) $t=3000$ for $D=0.02$ and $g=0.025$. As an initial condition, $n(x,y)$ is set to $n(i\Delta x,j\Delta y)=0.4+r(i\Delta x,j\Delta y)$, where $r$ is a random number between -0.01 and 0.01, and $\phi=0$.  
The uniform state is unstable and many small aggregates appear locally at $t=1000$. Local aggregates merge and larger aggregates appear at $t=3000$. Branching structures are observed for each aggregate.  In previous numerical simulations of the same model at different parameter values, no branching patterns were observed clearly, because the cell motion toward the cluster centers was rather fast~\cite{Sakaguchi2}. 

If the initial condition is set to $\phi=0$, $n(i\Delta x,j\Delta y)=0.4+r(i\Delta x,j\Delta y)$, where $r$ is a random number between -0.01 and 0.01, and $n(x,y)$ is set to 1 at $t=0$ in the central region $\sqrt{(x-L/2)^2+(y-L/2)^2}<0.5$, the cells tend to move toward the center because the initial concentration is high there. The local clusters of aggregation appear along the azimuthal direction owing to the instability. Figures 2(a) and 2(b) show two snapshots of regions satisfying $n>0.4$ at (a) $t=1000$ and (b) 3000. The parameter values are  the same as those in Fig.~1, i.e.,$\gamma=1,\nu=0.01,g=0.025,D=0.02$, and $c=0.0005$. Many notches of low concentration appear in the surroundings at $t=1000$. A branching structure appears at $t=3000$. This is related to the streaming instability discussed by H\"ofer and Maini~\cite{Maini}. The flow toward the center and local clustering in the azimuthal direction generate a branching pattern. 

We have performed numerical simulation using another initial condition under which $n(i\Delta x,j\Delta x)$ takes 0.1 or 0.9 randomly with a probability of 0.8 or 0.2, and $n(x,y)$ is set to 1 at $t=0$ in the central region $\sqrt{(x-L/2)^2+(y-L/2)^2}<0.5$. Figure 3 shows four snapshot patterns of regions satisfying $n>0.4$ at (a) $t=500$, (b) 750, (c) 2500, and (d) 10000.  The parameter values are the same as those in Figs.~1 and 2. Small aggregates appear locally at $t=500$. They tend to move toward the center. 
The small aggregates merge to form a branching pattern at $t=2500$. 
Cells move toward the center and branches shrink slowly.  
 Finally, a circular cluster appears at $t=10000$ as a result of the aggregation dynamics toward the center.  Branching patterns appear only in the aggregation process and the final state is the circular clustered state. This type of aggregation process is observed experimentally. As shown in Figs.~1-3, the aggregation process seems rather different if the initial conditions are different.   

We performed numerical simulations at different parameter values using the same initial condition as that in Fig.~2, that is,  $\phi=0$, $n(x,y)=1$ in the central region, and $n(x,y)=0.4+r(x,y)$, where $r(x,y)$ is a random number between -0.01 and 0.01 in the outer region.
Figures 4(a)-4(c)  show snapshots of regions satisfying $n>0.4$ (a) at $t=3000$ for $D=0.01$, (b) at $t=5000$ for $D=0.06$, and (c) at $t=10000$ for $D=0.2$. The other parameter values are $\gamma=1$, $\nu=0.01$, $g=0.025$, and $c=0.0005$. 
The branches at $D=0.01$ shown in Fig.~4(a) are thinner and denser than those at $D=0.02$ shown in Fig.~2(b). The branches become thicker as $D$ increases, and change into a rough interface of $n=0.4$ in Fig.~4(b) at $D=0.06$. The interface becomes smooth at $D=0.2$ in Fig.~4(c). For $D\ge 0.08$, the patterns shrink smoothly with time and no branching patterns are observed. Figure 2(d) shows a transition line in the parameter space of $(g,D)$ for $\gamma=1,\nu=0.01$, and $c=0.0005$ above which no branching patterns are observed.

\begin{figure}
\begin{center}
\includegraphics[height=3.5cm]{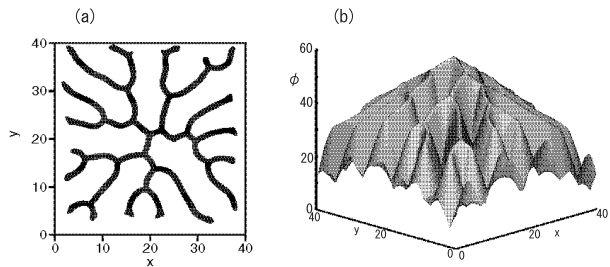}
\end{center}
\caption{(a) Branching pattern of $n$  and (b) 3D plot of $\phi$ at $t=75000$ for $f=0.000005$.}
\label{f5}
\end{figure}
\section{Stationary Branching Patterns}
To understand the mechanism of branching, it is desirable that a branching pattern appears in a stationary state. If a small number of cells are uniformly and constantly supplied in the whole region, a branching structure might be maintained. Since the total number of cells increases indefinitely by a constant supply, some cells should be removed. We have performed numerical simulation of another phenomenological model in which a constant $f$ is added on the right-hand side of Eq.~(2): 
\begin{eqnarray}  
\frac{\partial\phi}{\partial t}&=&\frac{n}{1+\gamma n}+\nu\nabla^2 \phi+g(\nabla \phi)^2,\\
\frac{\partial n}{\partial t}&=&D\nabla^2n-c\left \{\frac{\partial}{\partial x}\left (\frac{\partial \phi}{\partial x}n\right )+\frac{\partial}{\partial y}\left (\frac{\partial \phi}{\partial y}n\right )\right \}+f. \label{2dmodel2}
\end{eqnarray} 
Furthermore, $n(x,y)$ is  set to $n_0$ when the cell concentration goes beyond a threshold $n_0$, which corresponds to a removal process of cells.  Figure 5(a) shows regions satisfying $n>0.4$ at $t=75000$ for $f=0.000005$ and $n_0=5$. Other parameters and the initial condition are the same as those in Fig.~3. The branching structure is maintained and does not evolve to a circular cluster as shown in Fig.~3(d). Figure~5(b) shows a 3D plot of $\phi$ at the same time. A mountainlike profile with many valleys and ridges appears. The branching pattern of $n$ corresponds to the ridge pattern of the mountainlike structure of $\phi$. 
Cells tend to aggregate into local clusters and move toward regions of higher $\phi$ along the ridges. As a result, a branching pattern is created.  
The branching pattern is interpreted as a kind of dense-branching morphology~\cite{Vicsek}.
It is similar to branching patterns in river network models~\cite{Shei, Takayasu}. The parameter $f$ corresponds to the constant rainfall onto the whole region in the river network model. However, water flow occurs along the valleys in the river networks. Our branching pattern is also related to branching patterns in a flow field~\cite{Shimokawa} and barrier discharge~\cite{Sakaguchi3}. In all these systems, the flow toward the center and local clustering in the azimuthal direction are important for the formation of branching patterns.

\end{document}